\documentclass[aps,prx,twocolumn,reprint,amsmath,amssymb,superscriptaddress,floatfix,footinbib,longbibliography, tikz, dvipsnames]{revtex4-1}
\usepackage{natbib}

\usepackage{amsmath}
\usepackage{amssymb}
\usepackage{amsthm}
\usepackage{physics}

\usepackage{array}
\usepackage{multirow}
\usepackage{enumerate}
\usepackage{bm}
\usepackage{cancel}

\usepackage{graphicx}
\usepackage{empheq}
\usepackage{float}
\usepackage{tabularx}

\newcolumntype{C}{>{\centering}X}

\usepackage{standalone}
\usepackage{tikz}
\usepackage{pgfplots}
\pgfplotsset{compat=1.5}
\usetikzlibrary{arrows}
\usetikzlibrary{decorations.markings}   
\usetikzlibrary{patterns}
\usetikzlibrary{decorations.pathmorphing}

\usepackage{xcolor}
\usepackage{url}
\usepackage[colorlinks]{hyperref}
\hypersetup{%
        plainpages=true,
        breaklinks=true,
        hypertexnames=false,
        pageanchor=true,
        colorlinks=true,
        linkcolor={MidnightBlue},
        citecolor={RedOrange},
        urlcolor={MidnightBlue},
        anchorcolor={black}
      }
\definecolor{python_green}{RGB}{43,160,43}
\definecolor{python_blue}{RGB}{32,119,180}
\definecolor{python_orange}{RGB}{255,127,15}
\usepackage[all]{hypcap} 

\usepackage{mleftright} 

\newcommand{\figref}[1]{\mbox{Fig.~\ref{#1}}}

\newcommand{\secref}[1]{\mbox{Sec.~\ref{#1}}}

\newcommand{\appref}[1]{\mbox{Appendix~\ref{#1}}} 
\renewcommand{\eqref}[1]{\mbox{Eq.~(\ref{#1})}}
\newcommand{\eqsref}[2]{\mbox{Eqs.~(\ref{#1})--(\ref{#2})}}
\newcommand{\secsref}[2]{\mbox{Secs.~\ref{#1}--\ref{#2}}}

\newcommand{\figpanel}[2]{Fig.~\hyperref[#1]{\ref*{#1}(#2)}}
\newcommand{\figpanels}[3]{Fig.~\hyperref[#1]{\ref*{#1}(#2)--(#3)}}
\newcommand{\figpanelNoPrefix}[2]{\hyperref[#1]{\ref*{#1}(#2)}}

\newcommand{\Hc}{\text{H.c.}}
\DeclareMathOperator{\arccosh}{arccosh} 
\newcommand{\R}{\mathbb{R}}

\newcommand{\Hint}{H_\text{int}}
\newcommand{\bathfreq}{\omega_B}
\newcommand{\atomfreq}{\omega_1}
\newcommand{\detuning}{\Delta}
\newcommand{\dist}{d}

\AtBeginDocument{%
    \newwrite\bibnotes
    \def\bibnotesext{Notes.bib}
    \immediate\openout\bibnotes=\jobname\bibnotesext
    \immediate\write\bibnotes{@CONTROL{REVTEX41Control}}
    \immediate\write\bibnotes{@CONTROL{%
    apsrev41Control,author="08",editor="1",pages="0",title="0",year="1"}}
     \if@filesw
     \immediate\write\@auxout{\string\citation{apsrev41Control}}%
    \fi
}%

\begin{document}

\title{Interaction between giant atoms in a one-dimensional structured environment}
\date{\today}

\author{Ariadna Soro}
\email{soro@chalmers.se}
\affiliation{Department of Microtechnology and Nanoscience, Chalmers University of Technology, 412 96 Gothenburg, Sweden}

\author{Carlos S\'{a}nchez Mu\~{n}oz}
\email{carlossmwolff@gmail.com}
\affiliation{Departamento de F\'{i}sica Te\'{o}rica de la Materia Condensada and Condensed Matter Physics Center (IFIMAC), Universidad Aut\'{o}noma de Madrid, Madrid, Spain}

\author{Anton Frisk Kockum}
\email{anton.frisk.kockum@chalmers.se}
\affiliation{Department of Microtechnology and Nanoscience, Chalmers University of Technology, 412 96 Gothenburg, Sweden}


\begin{abstract}

Giant atoms---quantum emitters that couple to light at multiple discrete points---are emerging as a new paradigm in quantum optics thanks to their many promising properties, such as decoherence-free interaction.
While most previous work has considered giant atoms coupled to open continuous waveguides or a single giant atom coupled to a structured bath, here we study the \emph{interaction} between \emph{two} giant atoms mediated by a structured waveguide, e.g., a photonic crystal waveguide.
This environment is characterized by a finite energy band and a band gap, which affect atomic dynamics beyond the Markovian regime.
Here we show that, inside the band, decoherence-free interaction is possible for different atom-cavity detunings, but is  degraded from the continuous-waveguide case by time delay and other non-Markovian effects.
Outside the band, where atoms interact through the overlap of bound states, we find that giant atoms can interact more strongly and over longer distances than small atoms for some parameters---for instance, when restricting the maximum coupling strength achievable per coupling point.
The results presented here may find applications in quantum simulation and quantum gate implementation.

\end{abstract}

\maketitle


\section{Introduction}

Harnessing atom-photon interactions is crucial in the fast developing field of waveguide quantum electrodynamics, both at the fundamental level, and for the potential applications in quantum computation and quantum simulation of many-body physics~\cite{Roy2017, Noh2017, Sheremet2021}.
An increasingly popular approach to study such interactions consists of coupling quantum emitters to structured baths, which have a distinctive energy spectrum with finite bands and band gaps.
In such a setup, atom-photon bound states are formed in the band gaps, where photons become exponentially localized in the vicinity of the atoms, inhibiting their decay~\cite{Douglas2015, Gonzalez-Tudela2015, Calajo2016, Chang2018}.
Even at the band edge of the continuum of propagating modes, atoms show fractional decay due to the influence of bound states~\cite{Bykov1975, John1994, Kofman1994}.
Furthermore, multiple atoms coupled to the same reservoir can interact through the overlap of their bound-state photonic wavefunctions~\cite{Bay1997, Lambropoulos2000, Shahmoon2013}.
These interactions can be tuned by modifying the frequencies of the atoms and their coupling strengths to the bath, which opens doors for applications in quantum simulation and computation~\cite{Scigliuzzo2021}.

Many different platforms have been used to demonstrate phenomena arising from the interaction between an atom and a structured environment~\cite{Carusotto2020}. These include cold atoms coupled to either photonic crystal waveguides~\cite{Hood2016} or to an optical lattice~\cite{Krinner2018, Stewart2020}, as well as superconducting qubits coupled to either a microwave photonic crystal~\cite{Liu2017, Sundaresan2019, Harrington2019} or to a superconducting metamaterial~\cite{Mirhosseini2018, Kim2021, Ferreira2021, Scigliuzzo2021}. 

A remarkable feature of superconducting qubits is that they behave as giant atoms when coupled to a waveguide at multiple discrete points, as demonstrated in recent experiments~\cite{Kannan2020a, Vadiraj2021}.
We typically refer to \textit{giant atoms}~\cite{FriskKockum2021} (GAs) as those that break the dipole approximation and therefore cannot be considered \textit{small} in comparison to the wavelength of the electromagnetic field they interact with.
Such atoms exhibit striking phenomena that include frequency-dependent decay rates and Lamb shifts~\cite{FriskKockum2014}, waveguide-mediated decoherence-free interaction~\cite{FriskKockum2018a, Carollo2020, Soro2022}, and oscillating bound states~\cite{Guo2020}.

Although it has been theorized that giant atoms exhibit interesting new physics in many different architectures~\cite{Gonzalez-Tudela2019, Du2022}, so far most experimental studies have focused on GAs coupled to surface acoustic waves~\cite{Gustafsson14, Aref2016, Manenti2017, Noguchi2017, Satzinger2018, Moores2018, Bolgar2018, Sletten2019, Bienfait2019, Andersson2019, Bienfait2020, Andersson2020} and microwave waveguides~\cite{Kannan2020a, Vadiraj2021}.
It is therefore natural to ask whether there is an advantage, with respect to small atoms, in coupling giant atoms to structured environments. 
While some recent works have touched upon this question~\cite{Longhi2020, Zhao2020, Wang21, Yu2021, Vega2021, Wang2021, Xiao2021, Cheng2021, Zhang2022}, most of them focused on just a single GA.

Here, we focus instead on the interaction between two GAs in a structured environment.
We model the structured environment as a one-dimensional array of cavities with nearest-neighbor interaction and study the behavior of GAs in the single-excitation regime.
We use both numerical simulations and complex-analysis techniques---resolvent formalism---to examine the dynamics of the emitters tuned to different regions of the band structure. 

Within the band, we show how decoherence-free interaction (DFI), i.e., the ability that `braided' giant atoms have to interact without relaxing into the waveguide~\cite{FriskKockum2018a}, can be deteriorated due to non-Markovian decay and time delay.
We also establish the values of atom-cavity detuning and spacing between the coupling points at which DFI is possible.

In the band gap, we show that GAs can interact through the overlap of bound states and we compare their interaction mechanism to that of small atoms.
We find that, for some parameters, giant atoms can interact more strongly and over longer distances than small atoms.
In particular, we conclude that preference of a giant-atom design over a small-atom design might depend on the experimental constraints and the intended application. 
Possible applications include quantum simulation~\cite{Zhang2022a, Daley2022}, as well as implementation of entangling or SWAP gates for quantum computing~\cite{Scigliuzzo2021}.

This article is structured as follows.
In \secref{sec:theory1}, we present a theoretical model of a single GA coupled to a one-dimensional structured bath at an arbitrary number of coupling points.
We employ the model to describe the dynamics of such a system in \secref{sec:dynamics}, both through the resolvent formalism and numerical methods.
In \secref{sec:theory2}, we expand the model to consider two giant atoms, each with two coupling points to the structured bath. 
We analyze and discuss the interaction between the GAs when they are tuned to the band in \secref{sec:inside} and outside the band in \secref{sec:outside}.
In particular, in \secref{sec:giant_vs_small}, we compare the strength and fidelity of the interaction between small atoms against giant atoms, both inside and outside the band. 
We conclude in \secref{sec:conclusion} with a summary and an outlook.
We also include two appendices with more detailed derivations of the numerical methods used for the simulations (\appref{app:Dynamics}) and the resolvent formalism used to derive the self-energy of the atoms and the probability amplitudes of the atomic states (\appref{app:SelfEnergy}).


\section{A single giant atom}
\label{sec:1GA}

\subsection{Theoretical model}
\label{sec:theory1}

\begin{figure}[t]
    \centering
    \includegraphics[width=\columnwidth]{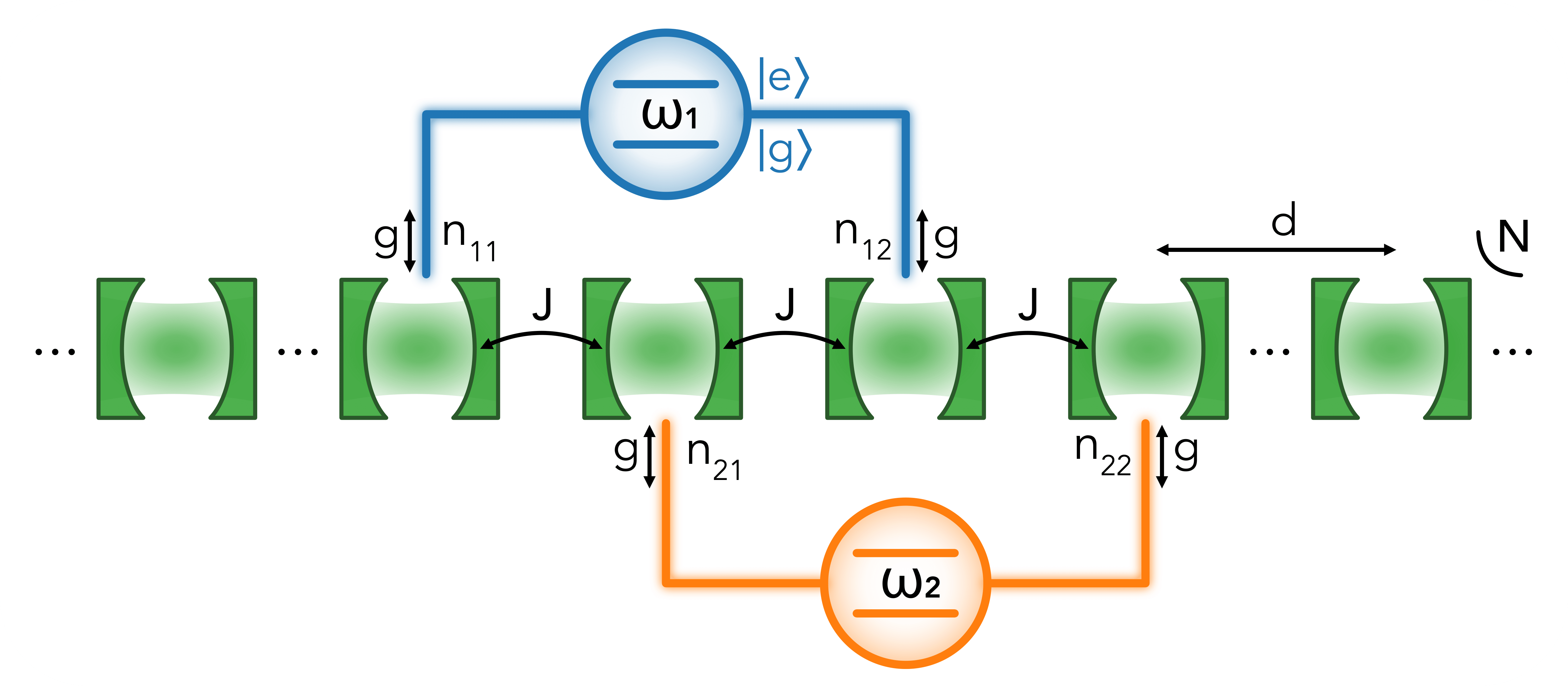}
    \caption{Two giant atoms in a braided configuration coupled to a 1D structured bath. The bath is modeled as an array of $N$ cavities with resonance frequency $\bathfreq$ and nearest-neighbor coupling strength $J$. The cavities are indexed by $n\in[0, N-1]$ (from left to right) and spaced by a unit length. We use $d$ as a measure of distance, which refers to the number of cavities spanned in that spacing. The atoms are two-level systems, with transition frequencies $\omega_i$ detuned from the middle of the band by $\detuning_i=\omega_i-\bathfreq$ ($i=1,2$). They are coupled to the cavities with coupling strength $g$ at each coupling point. We label each coupling point by $ip$, where $i$ refers to the atom and $p$ to the connection point, starting with $p=1$ as the leftmost one. The cavity to which the coupling point $ip$ is connected is denoted by $n_{ip}$ (or $n_p$ in case there is only one atom).}
    \label{fig:setup}
\end{figure}

The theoretical treatment presented here is similar to that introduced in Refs.~\cite{Cohen-Tannoudji, Gonzalez-Tudela2017, Wang21, Wang2021}, modified to describe a giant atom coupled to a one-dimensional (1D) structured environment (see \figref{fig:setup}).

The structured reservoir can be described as an array of $N$ cavities of frequency $\bathfreq$ with nearest-neighbor coupling $J$.
By taking their separation as the unit of length, we can characterize the position of the cavities with integer index $n\in[0,N-1]$, and thus label each corresponding cavity annihilation operator as $a_n$. The resulting bath Hamiltonian in real space, rotating at frequency $\bathfreq$, reads ($\hbar = 1$ throughout this article):
\begin{equation}
    H_B = -J\sum_{\expval{n,m}} (a_n^\dagger a_m + \Hc),
     \label{eq:H_B_real}
\end{equation}
where $\expval{n,m}$ denotes summation over all neighboring cavities $n$ and $m$, and H.c.~denotes Hermitian conjugate.

The Hamiltonian can be diagonalized by introducing periodic boundary conditions and the operators 
\begin{equation}
a_n=\frac{1}{\sqrt{N}}\sum_k a_k e^{-ikn},
\end{equation}
with $k\in [-\pi, \dots, \pi-\frac{2\pi}{N}]$. In that basis,
\begin{equation}
        H_B = \sum_k \omega(k)a_k^\dagger a_k,
        \label{eq:H_B_momentum}
\end{equation}
with $\omega(k) = -2J\cos(k)$.
The energy dispersion is linear around the middle of the band and parabolic close to the band edges, which translates into a density of states
\begin{equation}
    D(E) = \frac{1}{\pi\sqrt{4J^2-E^2}}\Theta(2J-\abs{E}),
    \label{eq:DOS}
\end{equation}
that is nearly constant around the middle of the band (i.e., for energies $E\approx 0$) and diverges at the band edges ($\abs{E/J}\approx 2$)~\cite{Gonzalez-Tudela2017}.

We now consider a two-level giant atom, with transition frequency $\atomfreq$, detuned from the middle of the band by $\detuning=\atomfreq-\bathfreq$. Then, its Hamiltonian is given by
\begin{equation}
    H_A = \detuning \sigma^+ \sigma^-,
    \label{eq:H_A}
\end{equation}
where $\sigma^\pm$ denote the atomic ladder operators.
If the giant atom couples to the bath at $P$ points, then their interaction can be described by
\begin{equation}
\begin{aligned}
    \Hint &= g  \sum_{p=1}^P \mleft(a_{n_p} \sigma^+ + \Hc \mright) \\
    &=\frac{g}{\sqrt{N}} \sum_k \sum_{p=1}^P \mleft(e^{-ikn_p} a_k\sigma^+ + \Hc \mright),
    \label{eq:H_int}
\end{aligned}
\end{equation}
where $n_p$ denotes the position of the bath mode which interacts with the $p$th coupling point.
Note that we have applied the rotating-wave approximation (RWA), which requires that $\atomfreq, \bathfreq\gg g$, a safe assumption to make in the optical and microwave regimes.

Finally, the total Hamiltonian of a single giant atom coupled to a structured reservoir at $P$ points is given by
\begin{equation}
H=H_B + H_A + \Hint
\label{eq:H}
\end{equation}
with the definitions above.

Using this model, we work in the single-excitation regime and assume coupling to a single polarization of light and a single bosonic band.
We focus on the predictions in the continuum limit, i.e., $N\to\infty$, where we can neglect the effects that arise from the finite size of the bath. 
We also neglect couplings to other reservoirs, by assuming the losses induced by such other couplings occur on a much longer time scale than the phenomena we want to study.
Lastly, we provide formulas for the regime $g\ll J$, although we illustrate the results for $g\lesssim J$.

We note that this theoretical model may be used to represent cold atoms coupled to photonic crystal waveguides~\cite{Hood2016} or optical lattices~\cite{Krinner2018, Stewart2020}, as well as superconducting qubits coupled to microwave photonic crystals~\cite{Liu2017, Sundaresan2019, Harrington2019} or superconducting metamaterials~\cite{Mirhosseini2018, Kim2021, Ferreira2021, Scigliuzzo2021}.
That being said, the experimental realization of cold atoms in the giant-atom regime remains elusive and thus we consider our setup to be most readily implementable with superconducting qubits.


\subsection{Dynamics}
\label{sec:dynamics}

In this Section, we  describe the methods used in this work to study the dynamics of GAs coupled to a structured bath, and we illustrate some key features in the case of a single giant atom. We show that, in the single-excitation subspace, a single GA with two coupling points spaced by a distance $\dist$ behaves in the same way as two small atoms~\cite{Gonzalez-Tudela2017} separated by the same distance, and we exemplify for $\dist=1,2$.
We also find that with an increasing number of coupling points, the decay rate of a GA as a function of detuning exhibits a more intricate profile.

One of the methods we employ to calculate the dynamics is direct numerical integration of the Schr\"{o}dinger equation.
While this method is easy to implement, it does not provide us information about the different contributions to the dynamics. 
Therefore, in order to dissect the dynamics and obtain a deeper understanding of the system, we also resort to the resolvent formalism.
The different approaches are presented hereunder.

\subsubsection{Energy spectrum}

\begin{figure}[t]
    \centering
    \includegraphics[width=\columnwidth]{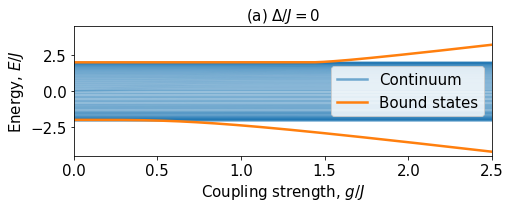}
    \includegraphics[width=\columnwidth]{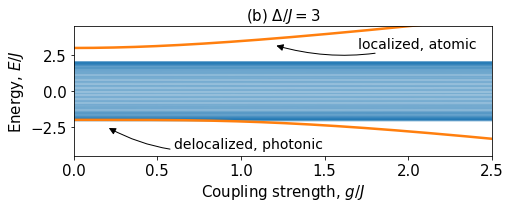}
    \caption{Energy spectra of a giant atom with two coupling points separated by a distance $\dist = 1$, as a function of the coupling strength $g/J$. (a) Atom-cavity detuning $\detuning/J = 0$. (b) Atom-cavity detuning $\detuning/J = 3$.}
    \label{fig:EnergySpectra}
\end{figure}

From the diagonalization of the full system-bath Hamiltonian $H$ [\eqref{eq:H}], we obtain the energy spectra illustrated in \figref{fig:EnergySpectra}.
In the spectra, we identify a continuum of states within the energies $E/J\in[-2,2]$, as given by the dispersion relation, and two bound states outside of the band, which is typical from a structured environment~\cite{John1990, John1991, Lambropoulos2000, Calajo2016}. Note that in the regime $g/J < 1$, the detuning $\detuning$ of the atom with respect to the middle of the band plays a big role, as will be illustrated by subsequent methods.
In particular, depending on the detuning and the coupling strength, the bound states can range from having a mainly atomic nature---localized and isolated from the continuum---to having a mainly photonic nature---delocalized and hybridized with the continuum---~\cite{Calajo2016, Scigliuzzo2021} (see \figref{fig:EnergySpectra}).

Hereafter, we will refer to the detunings in the range $\abs{\detuning/J} < 2$ as \emph{inside the band} or \emph{continuum of modes}, to $\abs{\detuning/J}=2$ as \emph{band edge}, and to $\abs{\detuning/J} > 2$ as \emph{outside the band} or \emph{band gap}.

\begin{figure*}[t]
    \centering
    \includegraphics[width=0.45\textwidth]{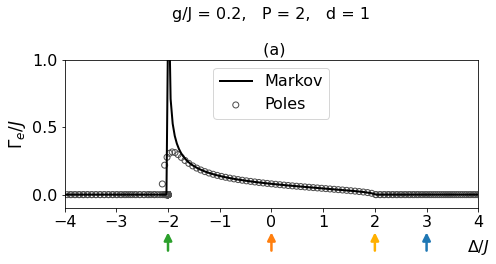}
    \includegraphics[width=0.45\textwidth]{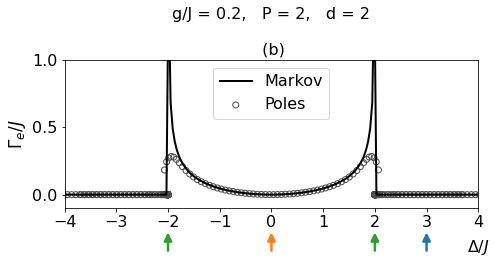}
    \includegraphics[width=0.45\textwidth]{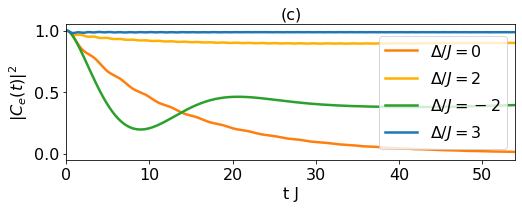}
    \includegraphics[width=0.45\textwidth]{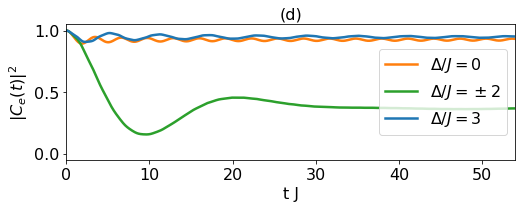}
    \caption{(a,b) Decay rate of a GA with $P=2$ coupling points separated by a distance (a) $\dist=1$, and (b) $\dist = 2$. The black solid line corresponds to the Markov prediction as described by \eqref{eq:Gamma_e}, while the gray markers are the poles of the Green's function [\eqref{eq:Green}]. (c,d) Atomic population of an initially excited GA with $P=2$ coupling points separated by a distance (c) $\dist = 1$, and (d) $\dist = 2$. The different lines show the dynamics for the atom detuned by $\detuning / J$ from the cavities, i.e. from the middle of the band. Those detunings are indicated with arrows of the corresponding color in (a,b).}
    \label{fig:Dynamics_GA12}
\end{figure*}

\begin{figure}[t]
    \centering
    \includegraphics[width=0.9\columnwidth]{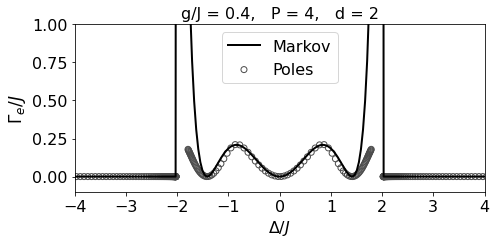}
    \caption{Decay rate of a GA with $P=4$ coupling points, each separated by $d=2$ cavities. The black solid line corresponds to the Markov prediction as described by \eqref{eq:Gamma_e},  while the gray markers are the poles of the Green's function [\eqref{eq:Green}].}
    \label{fig:DecayRate_P4}
\end{figure}


\subsubsection{Dynamics: resolvent formalism}
\label{sec:dynamics_analytical}

The probability amplitude of the atomic population of an initially excited GA can be analytically written for $t>0$ as~\cite{Cohen-Tannoudji, Gonzalez-Tudela2017, Wang21}:
\begin{equation}
    C_e(t) = -\frac{1}{2\pi i} \int_{-\infty}^\infty G_{e}(E + i 0^+) e^{-iEt}\,dE,
    \label{eq:C_e}
\end{equation}
which is the displaced Fourier transform of the retarded Green's function
\begin{equation}
    G_{e}(z) = \frac{1}{z-\detuning - \Sigma_{e}(z)}.
    \label{eq:Green}
\end{equation}
As detailed in \appref{app:SelfEnergy}, $G_{e}(z) = \mel{e}{G(z)}{e}$ is the matrix element of the resolvent of the Hamiltonian $G(z) = (z-H)^{-1}$, while $\Sigma_{e}(z)= \mel{e}{\Sigma(z)}{e}$ is the so-called self-energy of the atom, which is the matrix element of the level-shift operator $\Sigma(z)\approx H_\text{int}+H_\text{int}\sum_k\ketbra{k}(z-H_A-H_B)^{-1}H_\text{int}$. For a giant atom with $P$ coupling points, the self-energy reads, for $\Re{z}\gtrless 0$,
\begin{equation}
\begin{aligned}
    \Sigma_{e}(z) =& \pm \frac{g^2}{\sqrt{z^2-4J^2}} \\
    &\times \mleft[P+2\sum_{p=1}^{P-1} p\mleft(\frac{-z\pm\sqrt{z^2-4J^2}}{2J}\mright)^{(P-p)\dist\,} \mright].
    \label{eq:SelfEnergy1}
\end{aligned}
\end{equation}
By separating $\Sigma_{e}$ into its real and imaginary parts, we can identify the frequency-dependent Lamb shift $\Delta_{e}(E)\in \R$ and the decay rate $\Gamma_{e}(E)\in\R$ as follows~\cite{Cohen-Tannoudji, Gonzalez-Tudela2017}:
\begin{equation}
    \Sigma_{e}(E) = \Delta_{e}(E) - i\frac{\Gamma_{e}(E)}{2}.
    \label{eq:Gamma_e}
\end{equation}

We note that the model as we have presented it uses the RWA too early in its derivation to yield an accurate value of the Lamb shift. 
We therefore leave this quantity outside the scope of this article and refer the reader to Ref.~\cite{FriskKockum2014} for an in-depth discussion about frequency-dependent Lamb shifts in giant atoms.

In this section, we focus on the decay rate $\Gamma_{e}(E)$, which is proportional to the density of states from \eqref{eq:DOS}, i.e., nearly constant around the middle of the band and divergent at the band edges.
This U-shape profile, however, can be counteracted by the interference effects of a GA.
See, for instance, \figpanels{fig:Dynamics_GA12}{a}{b}, which displays the decay rate of a GA with $P=2$ coupling points and distances $\dist=1, 2$.
It is clear that the interference between the coupling points plays a crucial role, suppressing the atomic decay at one band edge ($\detuning/J=2$) for $\dist=1$ and at the band center ($\detuning/J=0$) for $\dist=2$.
As shown in Ref.~\cite{Gonzalez-Tudela2017}, the same destructive interference occurs between two small atoms in the single-excitation regime separated by the same distances.
In \figref{fig:DecayRate_P4}, we depict how with multiple coupling points (in this case, $P=4$), the suppression of the decay rate can occur for multiple detunings.

\begin{figure}[t]
    \centering
    \includegraphics[trim={5cm 0 5cm 0},clip, width=\columnwidth]{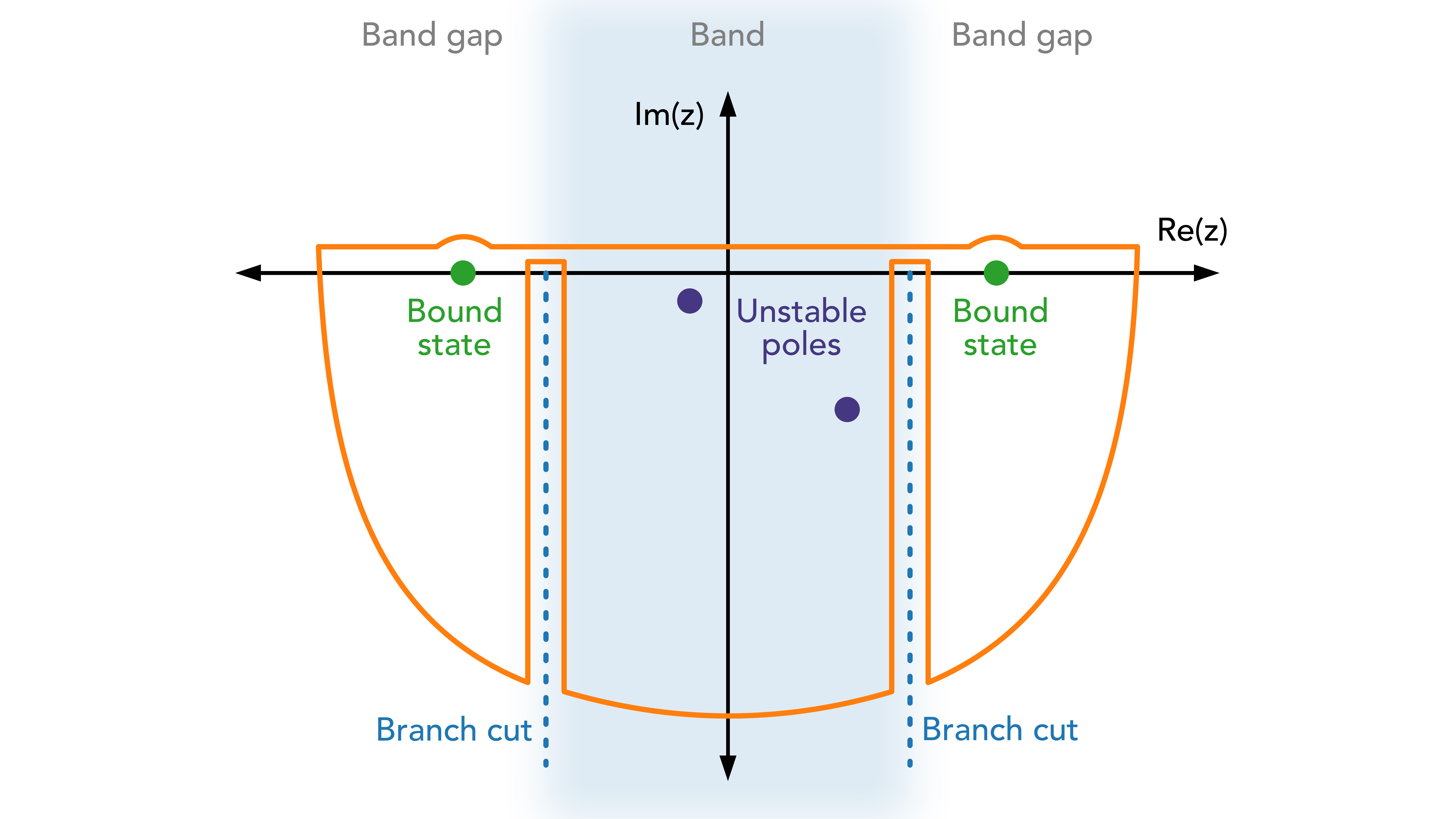}
    \caption{Contour of the integral in \eqref{eq:C_e}, with contributions from the poles of the Green's function [\eqref{eq:Green}] and the branch cuts at the band edges ($\abs{\Re(z)/J} = 2$).  Figure inspired by Ref.~\cite{Gonzalez-Tudela2017}.}
    \label{fig:ContourIntegral}
\end{figure}

We remark that the decay rate given by \eqref{eq:Gamma_e} and plotted as a continuous line in \figpanels{fig:Dynamics_GA12}{a}{b} is Markovian, and it does not account for the effect of the band edges.
In fact, the energy dispersion introduces branch cuts at the band edges, making the integral in \eqref{eq:C_e} contour around them (see \figref{fig:ContourIntegral}).
Essentially, this means that the atomic population $\abs{C_e(t)}^2$ [plotted in \figpanels{fig:Dynamics_GA12}{c}{d}] is affected by two elements: detours around the branch cuts, and poles of the Green's function [\eqref{eq:Green}].
In particular, complex or unstable poles with $\abs{\Re(z)/J}<2$ are responsible for the spontaneous emission into the bath, while real poles with $\abs{\Re(z)/J}>2$ are associated to the bound states.

As explained in Ref.~\cite{Gonzalez-Tudela2017}, the probability amplitude $C_e(t)$ can calculated as a sum of the different contributions:
\begin{equation}
    C_e(t) = \sum_{\alpha\in\substack{\text{branch}\\[-0.5pt]\text{cuts}}} C_{\alpha} (t) + \sum_{\beta \in \text{poles}} R_{\beta} e^{-iz_\beta t},
    \label{eq:contributions}
\end{equation}
where $R_\beta$ is be the residue of the real and unstable poles that we obtain through the residue theorem and that gives the overlap of the initial wave function with the poles, i.e.,
\begin{equation}
    R_\beta = \eval{\frac{1}{1-\partial_z \Sigma_{e}(z)}}_{z=z_\beta}.
\end{equation}

The poles provide a more accurate profile of the decay rate and highlight the non-Markovianity close to the band edges, as depicted in \figpanels{fig:Dynamics_GA12}{a}{b} and \figref{fig:DecayRate_P4} by gray markers. The non-Markovianity manifests as fractional atomic decay at the band edges~\cite{Bykov1975, John1994, Kofman1994}, as explained in the following \secref{sec:dynamics_numerical}.

We note that the resolvent formalism has been used before in structured environments to study small atoms~\cite{Gonzalez-Tudela2017} and chiral emission of giant atoms~\cite{Wang21, Wang2021}.
In this manuscript, we will use it instead to explain the interaction mechanism between two GAs, particularly to examine decoherence-free interaction (\secsref{sec:inside}{sec:outside}).

\subsubsection{Dynamics: numerical simulation}
\label{sec:dynamics_numerical}

Since solving \eqref{eq:C_e} analytically is not straightforward, it is sometimes convenient to simulate the dynamics numerically. For that, we use the spectral method~\cite{Press07, Gonzalez-Tudela2017}
to solve the Schr\"{o}dinger equation in discrete time steps $dt$:
\begin{equation}
    \ket{\psi(t_{n+1})} = U_{k\to n} e^{-iH_B dt} U_{n\to k} e^{-i(H_A + \Hint)dt} \ket{\psi(t_n)},
    \label{eq:spectral_method}
\end{equation}
where $t_n=n\,dt$ and $U_{n\to k}$ denotes the change between real ($n$) and momentum ($k$) space.
The key point of this method is that we can use the fact that $H_A+\Hint$ is just a $2\times2$ matrix in real space and that $H_B$ is diagonal in momentum space.
In particular, for each time step, we proceed as detailed in \appref{app:Dynamics}.

In \figpanels{fig:Dynamics_GA12}{c}{d}, we illustrate the dynamics of a giant atom  with $P=2$ coupling points separated by $\dist=1,2$.
We observe that, for $\dist=1$, around the center of the band ($\detuning/J=0$), the atom decays exponentially as is the case in a common transmission line or optical fiber.
For $\dist=2$, the decay is suppressed by destructive interference---in agreement with \figpanel{fig:Dynamics_GA12}{b}---, leading to a subradiant state~\cite{Dicke54, Lenz93, LeKien05, Lalumiere13}. This subradiance is also exhibited by two small atoms in the single-excitation regime, as reported in Ref.~\cite{Gonzalez-Tudela2017}.

When detuning the atom to the edges of the band ($\abs{\detuning/J}=2$), we observe fractional decay caused by a hybridization with the bound state, which can be seen in the spectra from \figref{fig:EnergySpectra}.
In \figpanel{fig:Dynamics_GA12}{a}, we note that for $\dist = 1$, $\Gamma_{e}$ at the lower band edge ($\detuning/J=-2$) is much larger than at the center of the band.
This translates into a faster initial decay, as is displayed in \figpanel{fig:Dynamics_GA12}{c}.
Nevertheless, the decay is soon counteracted by the hybridization with the bound state, which results in fractional decay.
At the upper band edge, decay is again suppressed by destructive interference between the coupling points of the atom.

Lastly, if the atom is well detuned from the band ($\abs{\detuning/J}=3$), it does not emit into the continuum, as the excitation is fully localized around the bound state (see \figref{fig:EnergySpectra}).
This behaviour is again equivalent to that of two small atoms in the single-excitation regime for $\dist=1,2$~\cite{Gonzalez-Tudela2017}.


\section{Two giant atoms}


\subsection{Theoretical model}
\label{sec:theory2}

Following the same approach as in \secref{sec:theory1}, we now consider the case of two giant atoms with $P=2$ coupling points each to the structured reservoir.
The full Hamiltonian is then
\begin{equation}
H=H_B + H_A + \Hint,
\end{equation}
where
\begin{align}
    H_B &= \sum_k \omega(k)a_k^\dagger a_k , \\
    H_A &= \detuning_1 \sigma_1^+ \sigma_1^- + \detuning_2 \sigma_2^+ \sigma_2^- , \\
    \Hint &= \frac{g}{\sqrt{N}} \sum_k \Big\{ \big[ \mleft(e^{ikn_{11}} + e^{ikn_{12}}\mright)\sigma_1^- \nonumber\\
    &\qquad\qquad\quad + \mleft(e^{ikn_{21}} + e^{ikn_{22}}\mright)\sigma_2^- \big]  a_k^\dagger + \Hc \Big\}.
\end{align}

As detailed in \appref{app:SelfEnergy}, two identical atoms (i.e., $\detuning_1=\detuning_2\equiv\detuning$ and $n_{12}-n_{11}=n_{22}-n_{21}\equiv d$) have a level-shift operator that is diagonal in the basis spanned by $\ket{\pm}=(\ket{eg}\pm\ket{ge})/\sqrt{2}$, with diagonal matrix elements $\Sigma_\pm (z) = \Sigma_e (z) \pm \Sigma_\text{int}(z)$.
Note that we now use the particular case of $\Sigma_e$ in \eqref{eq:SelfEnergy1} for $P=2$, i.e.,
\begin{align}
    \Sigma_e(z) &= \pm \frac{2 g^2}{\sqrt{z^2-4J^2}} \mleft[1 + f_\pm^d(z) \mright],
    \label{eq:Sigma_e_2}
\end{align}
where, for $\Re{z}\gtrless 0$,
\begin{equation}
    f_\pm(z) = \frac{-z\pm\sqrt{z^2-4J^2}}{2J}.
    \label{eq:f(z)}
\end{equation}
The interaction term reads, for $\Re{z}\gtrless 0$,
\begin{align}
    \Sigma_{\text{int}}(z) &= \pm \frac{g^2}{\sqrt{z^2-4J^2}}\nonumber\\
    &\quad\times\Big[f_\pm(z)^{\abs{n_{11}-n_{21}}} + f_\pm(z)^{\abs{n_{12}-n_{21}}} \nonumber\\
    &\qquad\;+ f_\pm(z)^{\abs{n_{11}-n_{22}}} + f_\pm(z)^{\abs{n_{12}-n_{22}}}\Big].
    \label{eq:Sigma_int}
\end{align}
Let us remark that the $\pm$ sign in \eqsref{eq:Sigma_e_2}{eq:Sigma_int} refers to $\Re{z}\gtrless 0$, whereas the $\pm$ sign outside of these expressions (e.g., in the definition of $\Sigma_\pm$) refers to the state $\ket{\pm}$.

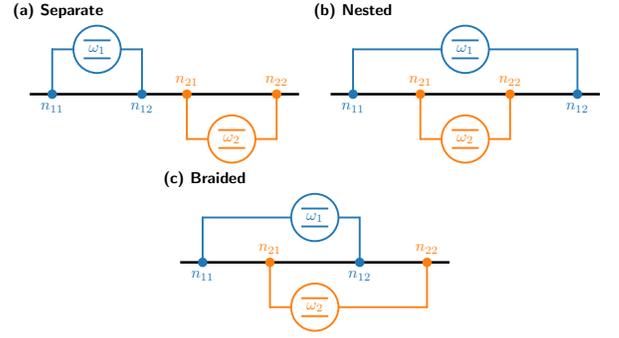
\begin{figure}[t]
\centering
\begin{tikzpicture}
\tikzset{every node}=[font=\normalsize\sffamily]

\node[anchor = north west] at (-3, 2.15) {\textbf{(a) Separate}};

\draw[ultra thick] (-2.5, 0) -- (3.5, 0);

\draw[very thick, python_blue] (-1, 1) circle (15pt);
\node[python_blue] at (-1, 1) {$\omega_1$};
\draw[very thick, python_blue] (-1.3, 1.2) -- (-0.7, 1.2);
\draw[very thick, python_blue] (-1.3, 0.8) -- (-0.7, 0.8);
\fill[python_blue] (-2, 0) circle (3pt);
\node[python_blue] at (-2, -0.3) {$n_{11}$};
\draw[very thick, python_blue] (-2, 0) -- (-2, 1.02);
\draw[very thick, python_blue] (-2.01, 1) -- (-1.53, 1);

\fill[python_blue] (0, 0) circle (3pt);
\node[python_blue] at (0, -0.3) {$n_{12}$};
\draw[very thick, python_blue] (0, 0) -- (0, 1.02);
\draw[very thick, python_blue] (0.01, 1) -- (-0.48, 1);

\draw[very thick, python_orange] (2,-1) circle (15pt);
\node[python_orange] at (2,-1) {$\omega_2$};
\draw[very thick, python_orange] (2.3, -1.2) -- (1.7, -1.2);
\draw[very thick, python_orange] (2.3, -0.8) -- (1.7, -0.8);
\fill[python_orange] (1, 0) circle (3pt);
\node[python_orange] at (1, 0.3) {$n_{21}$};
\draw[very thick, python_orange] (1, 0) -- (1, -1.02);
\draw[very thick, python_orange] (0.99, -1) -- (1.47, -1);

\fill[python_orange] (3, 0) circle (3pt);
\node[python_orange] at (3, 0.3) {$n_{22}$};
\draw[very thick, python_orange] (3, 0) -- (3, -1.02);
\draw[very thick, python_orange] (3.01, -1) -- (2.52, -1.0);

\end{tikzpicture}
\begin{tikzpicture}
\tikzset{every node}=[font=\normalsize\sffamily]


\node[anchor = north west] at (-3.5, 2.15) {\textbf{(b) Nested}};

\draw[ultra thick] (-3, 0) -- (3, 0);

\draw[very thick, python_blue] (0, 1) circle (15pt);
\node[python_blue] at (0, 1) {$\omega_1$};
\draw[very thick, python_blue] (-0.3, 1.2) -- (0.3, 1.2);
\draw[very thick, python_blue] (-0.3, 0.8) -- (0.3, 0.8);
\fill[python_blue] (-2.5, 0) circle (3pt);
\node[python_blue] at (-2.5, -0.3) {$n_{11}$};
\draw[very thick, python_blue] (-2.5, 0) -- (-2.5, 1.02);
\draw[very thick, python_blue] (-2.5, 1) -- (-0.53, 1);

\fill[python_blue] (2.5, 0) circle (3pt);
\node[python_blue] at (2.5, -0.3) {$n_{12}$};
\draw[very thick, python_blue] (2.5, 0) -- (2.5, 1.02);
\draw[very thick, python_blue] (2.51, 1) -- (0.52, 1.0);

\draw[very thick, python_orange] (0, -1) circle (15pt);
\node[python_orange] at (0, -1) {$\omega_2$};
\draw[very thick, python_orange] (-0.3, -1.2) -- (0.3, -1.2);
\draw[very thick, python_orange] (-0.3, -0.8) -- (0.3, -0.8);
\fill[python_orange] (-1, 0) circle (3pt);
\node[python_orange] at (-1, 0.3) {$n_{21}$};
\draw[very thick, python_orange] (-1, 0) -- (-1, -1.02);
\draw[very thick, python_orange] (-1.01, -1) -- (-0.52, -1);

\fill[python_orange] (1, 0) circle (3pt);
\node[python_orange] at (1, 0.3) {$n_{22}$};
\draw[very thick, python_orange] (1, 0) -- (1, -1.02);
\draw[very thick, python_orange] (0.99, -1) -- (0.52, -1);

\end{tikzpicture}
\begin{tikzpicture}
\tikzset{every node}=[font=\normalsize\sffamily]

\node[anchor = north west] at (-3.5, 2.15) {\textbf{(c) Braided}};

\draw[ultra thick] (-3, 0) -- (3, 0);

\draw[very thick, python_blue] (0, 1) circle (15pt);
\node[python_blue] at (0, 1) {$\omega_1$};
\draw[very thick, python_blue] (-0.3, 1.2) -- (0.3, 1.2);
\draw[very thick, python_blue] (-0.3, 0.8) -- (0.3, 0.8);
\fill[python_blue] (-2.5, 0) circle (3pt);
\node[python_blue] at (-2.5, -0.3) {$n_{11}$};
\draw[very thick, python_blue] (-2.5, 0) -- (-2.5, 1.02);
\draw[very thick, python_blue] (-2.5, 1) -- (-0.53, 1);

\fill[python_blue] (1, 0) circle (3pt);
\node[python_blue] at (1, -0.3) {$n_{12}$};
\draw[very thick, python_blue] (1, 0) -- (1, 1.02);
\draw[very thick, python_blue] (0.99, 1) -- (0.52, 1);

\draw[very thick, python_orange] (0, -1) circle (15pt);
\node[python_orange] at (0, -1) {$\omega_2$};
\draw[very thick, python_orange] (-0.3, -1.2) -- (0.3, -1.2);
\draw[very thick, python_orange] (-0.3, -0.8) -- (0.3, -0.8);
\fill[python_orange] (-1, 0) circle (3pt);
\node[python_orange] at (-1, 0.3) {$n_{21}$};
\draw[very thick, python_orange] (-1, 0) -- (-1, -1.02);
\draw[very thick, python_orange] (-1.01, -1) -- (-0.52, -1);

\fill[python_orange] (2.5, 0) circle (3pt);
\node[python_orange] at (2.5, 0.3) {$n_{22}$};
\draw[very thick, python_orange] (2.5, 0) -- (2.5, -1.02);
\draw[very thick, python_orange] (2.51, -1) -- (0.52, -1.0);

\end{tikzpicture}
\caption{Different arrangements of two giant atoms with two coupling points each. (a) Separate: $n_{11}<n_{12}<n_{21}<n_{22}$. (b) Nested: $n_{11}<n_{21}<n_{22}<n_{12}$. (c) Braided: $n_{11}<n_{21}<n_{12}<n_{22}$.}
\label{fig:giant_atoms}
\end{figure}

In particular, if all the coupling points are equidistant and separated by $\dist$, then the three possible topologies of giant atoms (see \figref{fig:giant_atoms}) have the following interaction term in the self-energy:
\begin{align}
    \Sigma_{\text{int}}^\text{sep}(z) &= \pm \frac{g^2}{\sqrt{z^2-4J^2}} \big[f_\pm(z)^{\dist} + 2 f_\pm(z)^{2\dist} \nonumber\\
    &\qquad\qquad\qquad\quad + f_\pm(z)^{3\dist} \big],\\
    \Sigma_{\text{int}}^\text{bra}(z) &= \pm \frac{g^2}{\sqrt{z^2-4J^2}} \mleft[3f_\pm(z)^{\dist} + f_\pm(z)^{3\dist} \mright],\label{eq:Sigma_int_bra}\\
    \Sigma_{\text{int}}^\text{nes}(z) &= \pm \frac{g^2}{\sqrt{z^2-4J^2}} \mleft[2f_\pm(z)^{\dist} + 2 f_\pm(z)^{2\dist} \mright].
\end{align}

In a similar fashion as in \eqref{eq:Gamma_e}, we note that $\Re{\Sigma_e}$ yields the atomic frequency shifts, $\Re{\Sigma_\text{int}}$ the exchange interaction between the atoms, $-2\Im{\Sigma_e}$ the individual decay rates and $-2\Im{\Sigma_\text{int}}$ the collective decay.

Finally, we find that the atomic probability amplitudes for the states $\ket{eg}$ and $\ket{ge}$ are (see \appref{app:SelfEnergy} for details):
\begin{equation}
C_{\substack{eg\\[-0.5pt]ge}}(t) = \frac{1}{2}[C_+(t)\pm C_-(t)],
\end{equation}
where $C_\pm(t)$ are the probability amplitudes corresponding to the states $\ket{\pm}$, with the Green's functions $G_\pm$ and the self-energies $\Sigma_\pm$:
\begin{equation}
C_\pm(t) =  -\frac{1}{2\pi i}\int_{-\infty}^\infty \frac{e^{-iEt} dE}{(E+i0^+) -\detuning-(\Sigma_{e}\pm\Sigma_{\text{int}})}.
\label{eq:C_pm}
\end{equation}


\subsection{Inside the band ($\abs{\detuning/J} < 2$)}
\label{sec:inside}


\subsubsection{Interaction mechanism}
\label{sec:interaction_inside}

One of the most interesting phenomena discovered in giant atoms is their ability to interact through a waveguide without decohering~\cite{FriskKockum2018a, Kannan2020a, Soro2022}.
When arranged in a braided configuration [see \figref{fig:setup} and \figpanel{fig:giant_atoms}{c}], destructive interference between the coupling points can suppress atomic relaxation---both individual and collective---, without cancelling the exchange interaction.
Therefore, two or more braided atoms can exchange an excitation back and forth, without losing it into the waveguide---thus the name decoherence-free interaction (DFI).
In this section, we show the differences between DFI mediated by a continuous waveguide and a structured waveguide, and we explain the mechanism behind it.

\begin{figure}[t]
    \centering
    \includegraphics[width=\columnwidth]{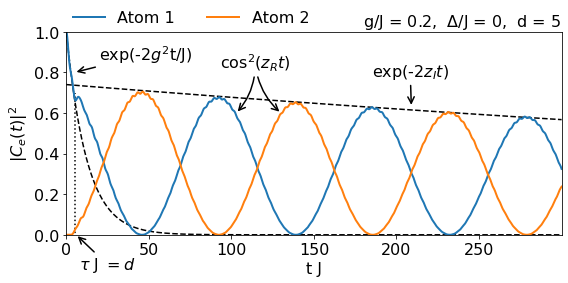}
    \caption{Decoherence-free interaction between two braided giant atoms with two coupling points each, spaced equidistantly by $d=5$ (i.e., the outermost points are separated by 15 cavities). Both atoms are tuned to the band center ($\detuning/J=0$) and coupled to the structured waveguide with coupling strength $g/J=0.2$. The parameters $z_R$ and $z_I$ are defined in \secref{sec:interaction_inside}.}
    \label{fig:DFI_inside}
\end{figure}

In \figref{fig:DFI_inside}, we show a numerical simulation of the population of two giant atoms in a braided configuration, according to the methods described in \secsref{sec:dynamics_analytical}{sec:dynamics_numerical} (both methods yield the same dynamics).
The atoms are laid out with subsequent coupling points separated by a distance $\dist=5$ and both atoms are tuned to the resonance frequency of the cavities, i.e., to the middle of the band.
We prepare the first atom in its excited state and the second in its ground state, and we then let the system evolve.

The dynamics in \figref{fig:DFI_inside} show the typical oscillations from DFI~\cite{FriskKockum2018a, Kannan2020a}, although they are far from the ideal case. 
In particular, there is first an exponential decay $e^{-2g^2t/J}$, which corresponds to the first atom spontaneously relaxing into the waveguide and not yet interacting with the second.
This decay occurs at twice the rate of the spontaneous emission of a small atom~\cite{Calajo2016, Gonzalez-Tudela2017} (since the giant atom couples at two points instead of one), and it is independent of the distance $\dist$ and the detuning $\detuning$.
Instead, it is the duration of such a decay what depends on the distance and the detuning:
it occurs for a time $\tau J = \dist$, which is the time the excitation takes to travel between the two coupling points of the first atom~\cite{Longhi2020, Du2022a, Du2022b}, and thus the time it takes to build up the necessary interference:
\begin{equation}
    \tau J = \frac{n_{12}-n_{11}}{v_g} J= \frac{2\dist}{2J} J= \dist,
\end{equation}
since the velocity $v_g(\detuning) = \eval{\partial_k \omega(k)}_{k=k(\detuning)} = \sqrt{4J^2-\detuning^2}$~\cite{Calajo2016, Gonzalez-Tudela2017}.

After this time, DFI begins and the atoms share the excitation back and forth.
These oscillations are caused by the real part of two unstable poles $z_+$ and $z_-$ of the Green's function $G_+$ and $G_-$, respectively.
In particular, the population evolves like $\cos^2(z_{R} t)$, where $z_{R} = \abs{\Re(z_+)-\Re(z_-)}/2$.
This can be explained through \eqref{eq:contributions}, where we see that, far from the branch cuts, the atomic probability amplitudes are a sum of exponential functions of the poles weighted by their residues.
In particular, DFI oscillations arise when two poles have similar residues, which implies $R_+e^{-iz_+t} + R_-e^{-iz_-t} \approx R_\pm\cos(z_Rt)$; and the absolute value of those residues is close to 1, meaning that there are no other major contributions to the dynamics. 
In the Markovian regime, i.e., when the poles coincide with the Markovian prediction [see \figpanels{fig:Dynamics_GA12}{a}{c} and  \figref{fig:DecayRate_P4}], then  $z_R=\Re[\Sigma_\text{int}(\detuning)]$ is the exchange interaction.
Unfortunately, these poles also have a small imaginary part, which causes an additional decay of the DFI oscillations $e^{-2 z_I t}$, where $z_I = \abs{\Im(z_+)+\Im(z_-)}/2$.

\begin{figure}[t]
    \centering
    \includegraphics[width=0.9\columnwidth]{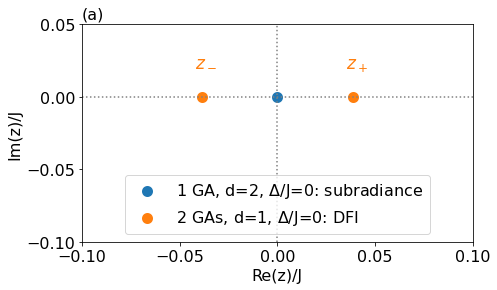}
    \includegraphics[width=0.86\columnwidth]{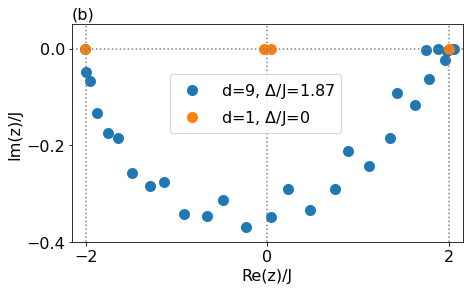}
    \caption{(a) Unstable poles of the Green's functions $G_{e}$ [\eqref{eq:Green}] (blue) and  $G_\pm$ (orange), for coupling strength $g/J=0.2$ and zero detuning. The blue pole is responsible for subradiance in a single giant atom with two coupling points spaced by $d=2$. The orange poles are responsible for decoherence-free interaction between two braided giant atoms with two coupling points each, spaced equidistantly by $\dist=1$. (b) Real and complex poles of the Green's functions $G_\pm$ of two braided giant atoms, for $\dist=9,\; \detuning/J=1.87$ (blue) and $\dist=1, \;\detuning/J=0$.}
    \label{fig:Poles_DFI_Subradiance}
\end{figure}

When $z_I\to 0$ (e.g., when $\detuning\to0$ and $d$ is small), we can think of DFI as a multiple-atom analogue of the subradiance observed in \figref{fig:Dynamics_GA12}.
The difference lies in the fact that, for multiple GAs, the symmetric (+) and antisymmetric ($-$) components give rise to two unstable poles, instead of one [see \figpanel{fig:Poles_DFI_Subradiance}{a}]. 
In turn, the poles cause the population exchanges between the atoms, i.e., the decoherence-free interaction.

As shown in \figpanel{fig:Poles_DFI_Subradiance}{b}, with increasing distance and detuning, more poles appear---which also occurs for a single GA---and, as their contributions become relevant, the DFI oscillations exhibit beatings and other anomalous behavior.
This can again be understood through \eqref{eq:contributions}, where the atomic probability amplitudes become a sum of exponential functions with different frequencies and weights.

\subsubsection{DFI points}
\label{sec:DFIpts}

After analyzing \figref{fig:DFI_inside}, it is natural to wonder how robust DFI is against time-delay effects, which in turn is affected by the separation between coupling points and the detuning from the cavity frequency.
In order to answer that question, it is imperative to first find where in the parameter space DFI is possible.  

\begin{figure}[t]
\raggedright
    \includegraphics[width=0.975\columnwidth]{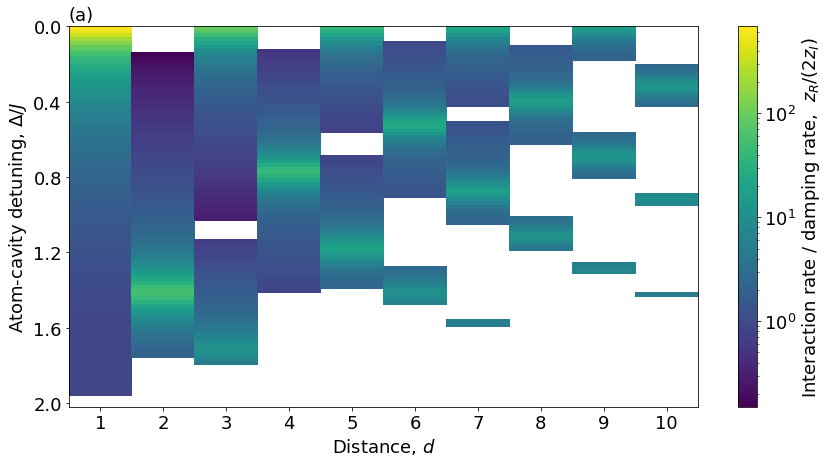}
    \includegraphics[width=\columnwidth]{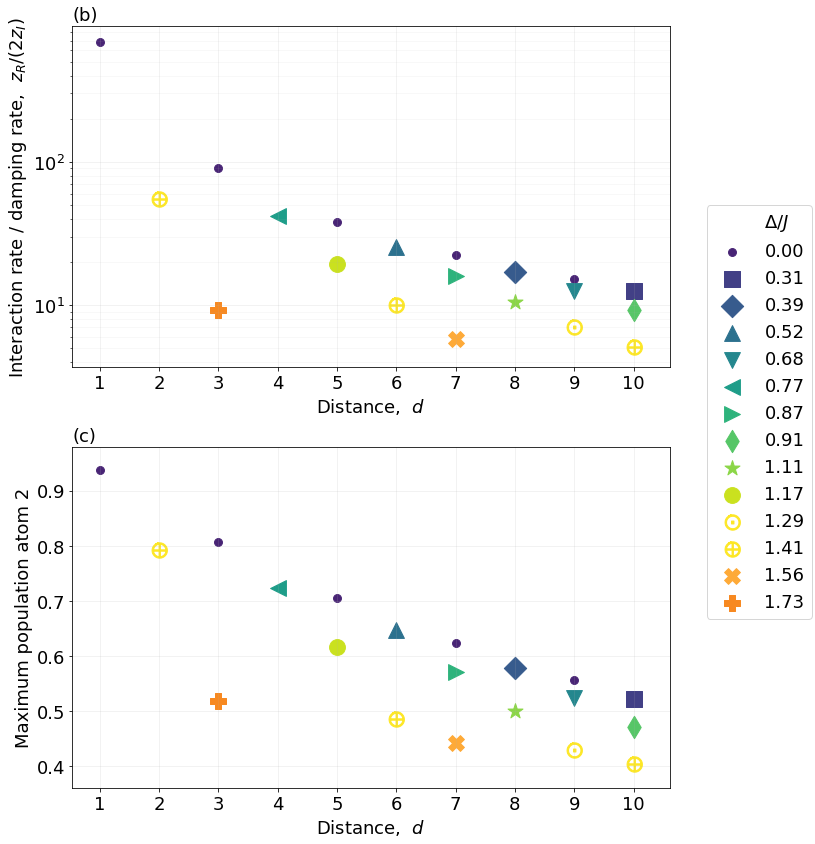}
    \caption{(a) Interaction rate over damping rate of DFI $z_R/(2z_I)$, plotted for atom-cavity detunings $\detuning/J$ inside the band and for distances between coupling points $\dist\le10$. (b) Maxima of the plot in (a), represented as interaction rate over damping rate of DFI $z_R/(2z_I)$, against distance $\dist$. (c) Maximum population transfer achievable in all the DFI points shown in (b).}
    \label{fig:DFIpts}
\end{figure}

We can calculate the possible DFI points by mapping the discrete variable $\dist\in\mathbb{N}$ to a continuous phase $\varphi\in[0,2\pi)$.
Inside the band, i.e., for $\detuning/J\in[-2,2]$, we find that $\abs{f(\detuning)}=1$ in \eqref{eq:f(z)}.
Therefore, we can write $f$ as a complex exponential $f = e^{i\phi}$, with $\phi = \arccos(-\detuning/2J)$.
Alternatively, we can write $f^d = e^{id\phi} = e^{i\varphi}$, with $\varphi=d\phi = d\arccos(-\detuning/2J)$.
This allows us to map the distances $\dist$ to the phase conditions for DFI, which are well established in the framework of a continuous waveguide~\cite{FriskKockum2018a, Soro2022, Carollo2020}.
In particular, we know that braided giant atoms with equidistant coupling points need to be spaced apart by $\varphi=\pi/2$ to interact without decohering.
There are infinitely many points that satisfy this condition with the mapping we have established.
Even if we restrict ourselves to a certain detuning, there will always be a distance that fulfills $\varphi=\pi/2$  ($\mkern-10mu\mod 2\pi$).
All the DFI points with $d\le 10$ (DFI is severely degraded for larger distances) are displayed in \figref{fig:DFIpts}.

This mapping from a discrete variable $d$ to a continuous one $\varphi$ only holds inside the band, where $\abs{f(\detuning)}=1$. Outside the band, $\abs{f(\detuning)}\neq1$ and we cannot convert it to a complex exponential.

A good figure of merit to quantify DFI is the ratio between the interaction rate---the frequency of the DFI oscillations---and the damping rate, $z_R/(2z_I)$.
In \figpanel{fig:DFIpts}{a}, we plot this ratio for different distances $d$ and different detunings $\detuning/J$, and we take the maxima as markers in \figpanel{fig:DFIpts}{b}. 
We note that we do not display points where $z_R/(2z_I)$ is ill-defined due to the presence of many unstable poles with relevant contributions, such as the case shown in \figpanel{fig:Poles_DFI_Subradiance}{b}.
Two aspects are clear immediately: DFI becomes worse with both increasing distance and increasing detuning from the band center.
We also note that the middle of the band is basically equivalent to a continuous waveguide, so structured environments in this context do not seem to provide any advantage.
That said, DFI can still occur at other energies within the band, which is a feature unique to braided giant atoms and unattainable for small atoms.

While \figpanel{fig:DFIpts}{b} quantifies the amount of population exchanges there can be between the two atoms, which can be useful to know for quantum-simulation applications,
it does not show any information about how much of the excitation is lost into the waveguide due to the first exponential decay.
This is instead depicted in \figpanel{fig:DFIpts}{c}, which represents the maximum population that the second atom reaches.
Essentially, this panel takes the fidelity of the population exchange as a different measure of good DFI, which proves more relevant in the implementation of quantum gates.
We note that the behavior of the two metrics is almost identical, including the fact that the continuous-waveguide case is better than all other detunings inside the band of the structured environment.
In conclusion, although the behavior of the two metrics is quite similar, in practice one might turn to one or the other depending on the purpose of the experimental implementation.

\subsection{Outside the band ($\abs{\detuning/J} > 2$)}
\label{sec:outside}

\begin{figure*}[t]
    \centering
    \includegraphics[width=\columnwidth]{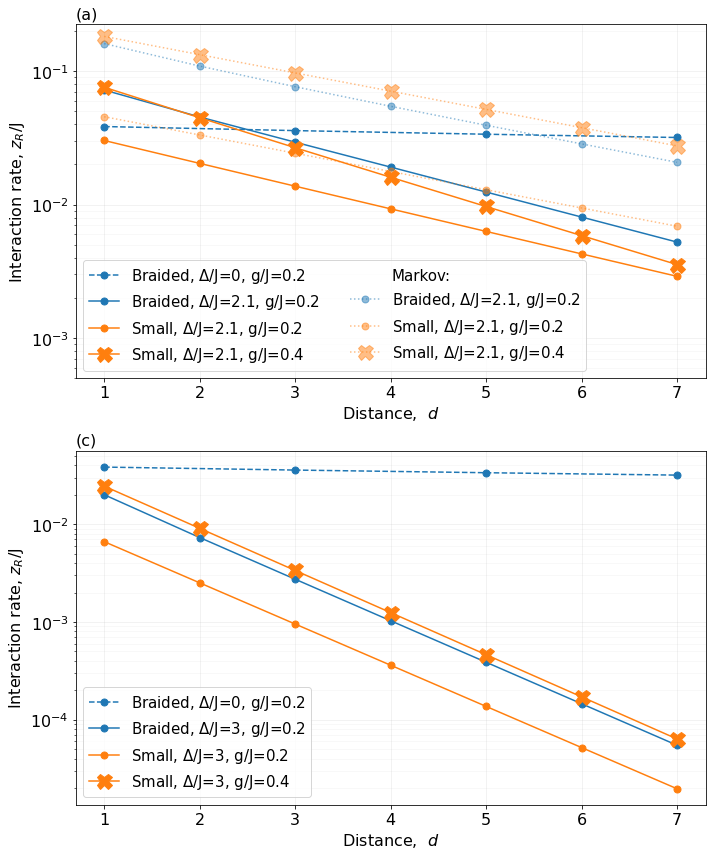}
    \includegraphics[width=\columnwidth]{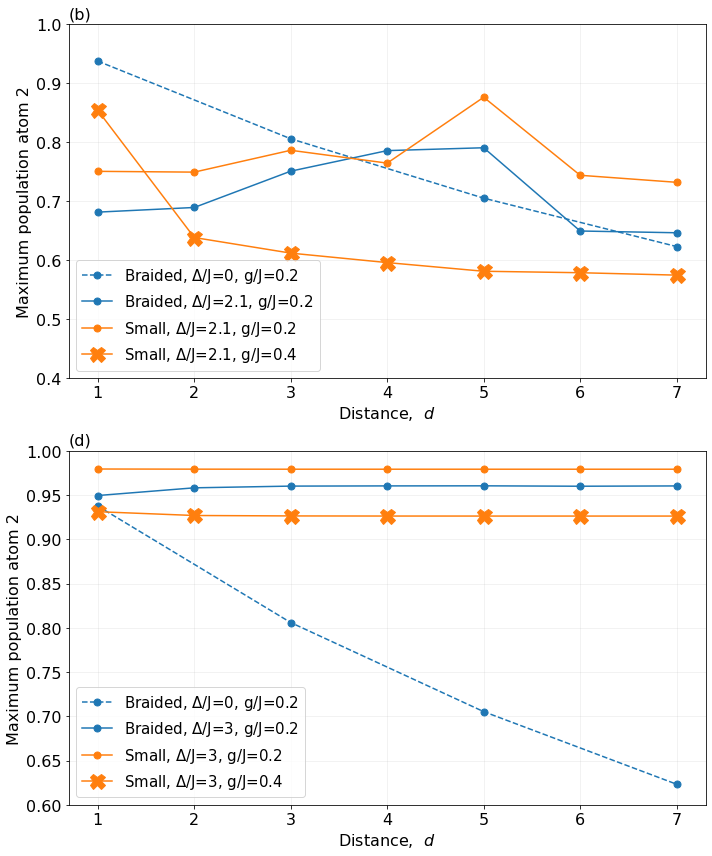}
    \caption{Comparison of interactions between small and giant atoms, inside and outside the band. We use color to distinguish between small (orange) and giant (blue) atoms, line style to differentiate between inside (dashed) and outside (solid) the band, and markers to distinguish different coupling strengths: $g/J=0.2$ (dots) and $g/J=0.4$ (crosses).
    \underline{Left:} Interaction rate $z_R$, plotted against distance between the coupling points $d$. Panel (a) shows the discrepancy between the interaction rate $z_R$ and the exchange interaction $\Re[\Sigma_\text{int}(\detuning)]$, while (c) corresponds to the Markovian regime, where $z_R=\Re[\Sigma_\text{int}(\detuning)]$ and the lines coincide. \underline{Right:} Maximum population of atom 2, plotted against distance between the coupling points $d$.
    \underline{Top:} Atoms in the band gap are tuned very close to the band edge, $\detuning/J=2.1$.
    \underline{Bottom:} Atoms in the band gap are tuned far from the band edge, $\detuning/J=3$.}
    \label{fig:Giant_vs_Small}
\end{figure*}


\subsubsection{Interaction mechanism}

When the atoms are tuned to the band gap, both small and giant atoms---in any configuration of the coupling points---can interact without decohering through the overlap of bound states~\cite{Bay1997, Lambropoulos2000, Shahmoon2013}.
The bound states are exponentially localized around the coupling points of the atoms, and therefore the atoms can only interact if they are close enough.
The interaction mechanism, in this case, differs from DFI inside the band at a physical level: there are no interference effects involved.

Let us compare the dynamics outside and inside the band by going back to \figref{fig:DFI_inside}.
The first exponential decay one can see there, corresponding to the buildup of interference, is not present when the atoms are tuned far outside the band.
This is because the atoms are decoupled from the continuum of modes, which can be understood as the coupling strength $g$ being effectively 0, thus resulting in $e^{-2g^2t/J} = 1$.
Only for detunings very close to the band edge ($\abs{\detuning/J}\gtrsim 2$) there can be some initial decay caused by the hybridization of the bound states---or, mathematically, by the contribution of the unstable poles and branch cuts.

Moreover, the interaction outside the band is mediated by the bound states, and therefore the main contribution to the dynamics are real poles of the Green's function.
Hence, $z_R\neq 0$ but $z_I=0$, which means the population exchanges are not damped over time, i.e., there is no decoherence.

Now, even though interaction through the bound states is fully decoherence-free and unaffected by time delay, it is not necessarily better than DFI inside the band.
The reason for that is that the interaction rate $z_R$ decays exponentially with the distance between coupling points, so the interaction becomes negligible at large $\dist$ \cite{Douglas2015, Gonzalez-Tudela2015, Chang2018, Wang21, Zhang2022a}.
This can be seen mathematically in the Markovian regime---e.g., far outside the band---, where the interaction rate equals the exchange interaction.
For instance, for braided atoms in the upper band gap,
\begin{align}
z_R =& \Re[\Sigma_\text{int}(\detuning)] \nonumber\\
=&\frac{g^2}{\sqrt{\detuning^2-4J^2}}\nonumber\\
&\times[ 3e^{-d\arccosh(\detuning/(2J))}+e^{-3d\arccosh(\detuning/(2J))}],    
\end{align}
where we used \eqref{eq:Sigma_int_bra} and that $\arccosh(x)=\ln(x+\sqrt{x^2-1})$.
As \figpanel{fig:Giant_vs_Small}{a} shows, this equality breaks down close to the band edge, although the exponential decay with distance remains.

\subsubsection{Giant versus small atoms}
\label{sec:giant_vs_small}

In this section, we analyze and compare the interaction between small atoms outside the band, and giant atoms inside and outside the band.
In particular, we show with different metrics that GAs can outperform small atoms in some regions of the parameter space, i.e., depending on the coupling strength, the distance between coupling points, and the atom-cavity detuning.

In the same way we defined two measures of good DFI inside the band in \secref{sec:DFIpts}, we can do it outside the band too.
We can either compare the performance of the atomic interaction in terms of strength, i.e., interaction rate [\figpanel{fig:Giant_vs_Small}{a} and \figpanelNoPrefix{fig:Giant_vs_Small}{c}], or in terms of the maximum population transfer [\figpanel{fig:Giant_vs_Small}{b} and \figpanelNoPrefix{fig:Giant_vs_Small}{d}]. Let us start with the former.

As we can see in the left-side plots of \figref{fig:Giant_vs_Small}, for the same coupling strength, braided giant atoms outperform small atoms both inside and outside the band for all distances and detunings.
However, as discussed in Ref.~\cite{Soro2022}, it might be unfair to compare giant and small atoms with the same coupling strength at each coupling point, since GAs have more coupling points.
Therefore, we also consider the case where the coupling strength of the small atoms is twice as much as the giant atoms, i.e., $g_{SA} = 2g_{GA}$.
In this case, small atoms perform almost identically as braided atoms outside the band except for very large distances and detuning close to the band edge, where they perform worse. 
It is important to note that, while the coupling points of the small atoms are separated by a distance $d$, the outermost points of the GAs are separated by three times as much ($3d$), meaning that GAs can actually keep the same interaction rate over longer distances. 

Similarly as in the previous section, \figpanel{fig:Giant_vs_Small}{a} and \figpanelNoPrefix{fig:Giant_vs_Small}{c} only account for the frequency of the population exchanges, so they do not include  the fact that the DFI oscillations are damped for braided atoms inside the band, or the fact that atoms tuned close to the band edge lose part of the excitation due to the hybridization of the bound states with the continuum of modes.
For these reasons, we also show the maximum population transfer plots in \figpanel{fig:Giant_vs_Small}{b} and \figpanelNoPrefix{fig:Giant_vs_Small}{d}.

Far outside the band, all atoms show population transfer over 90\%, which remain constant with distance.
However, tuning atoms close to the band edge significantly reduces that figure due to the effect of unstable poles and branch cuts. 
In this case, the maximum population transfer does not remain constant with distance.
On the contrary, it can be enhanced or reduced by the effect of the other poles.
Lastly, tuning braided atoms to the band center offers a population transfer ranging from over 90\% and decreasing exponentially with distance to around 60\% at $d=7$.

Overall, good population exchanges come at the cost of low interaction rates and vice versa. 
Therefore, it is a matter of finding a suitable trade-off between the two depending on the application: higher interaction rates are needed for fast quantum gates, whereas slow but good population transfer is more suitable for quantum simulation and adiabatic processes.

To conclude, as mentioned before, it is not straightforward to compare giant against small atoms, nor interactions inside versus outside the band.
However, at a practical level, experimental constraints sometimes impose a limit on the maximum coupling strength achievable in one coupling point, so if that is the case, then one might want to consider a braided giant-atom design to improve interaction.
Furthermore, until now, we have neglected external sources of decoherence such as coupling to other baths, but in experiments, these sources might become unavoidable and non-negligible when the interaction rates are low, e.g., outside of the band. In this context, braided giant atoms tuned to the band might also be preferable over small atoms to improve interaction.


\section{Conclusion}
\label{sec:conclusion}

In this manuscript, we have characterized the coupling of a giant atom to a one-dimensional structured environment, as well as the interaction mechanism between two giant atoms, both analytically and numerically.

In particular, with the atoms tuned to the band, we have shown that decoherence-free interaction is best in the continuous-waveguide case, but possible for other detunings.
We have also demonstrated, through different metrics, how decoherence-free interaction deteriorates exponentially with increasing distance between the coupling points.
Using resolvent-formalism techniques, we have dissected the dynamics into different contributions and shown the significance of time delay and other non-Markovian effects.
Lastly, we identified decoherence-free interaction as the multiple-giant-atom analogue of subradiance.

Furthermore, we have presented an extensive comparison between giant and small atoms---inside and outside the band---to determine which of them interact at a higher rate, with higher fidelity and over longer distances.
We have concluded that the answer depends on three parameters: the coupling strength, the distance between coupling points and the detuning of the atoms from the cavities.
In particular, giant atoms can provide an advantage over small atoms in some regions of the parameter space, for instance, when restricting the maximum coupling strength achievable per coupling point.
We have also found that there is a trade-off between good population exchanges and high interaction rates, and that the choice of parameters may depend on the application: higher interaction rates are needed for fast quantum gates, whereas slow but good population transfer is more suitable for quantum simulation and adiabatic processes.

With the technology available today, the model presented here is readily implementable with superconducting qubits coupled to either a microwave photonic crystal~\cite{Liu2017, Sundaresan2019, Harrington2019} or to a superconducting metamaterial~\cite{Mirhosseini2018, Kim2021, Ferreira2021, Scigliuzzo2021}.
Moreover, as it has been showed in recent experiments, this platform can be used for quantum simulation of spinless bosonic models~\cite{Zhang2022, Daley2022} and for the implementation of entangling or SWAP gates~\cite{Kannan2020a, Scigliuzzo2021}.

Future work may include extending the model to higher-dimensional structured environments~\cite{Gonzalez-Tudela2017, Gonzalez-Tudela2019} and to more elaborate band structures~\cite{Carusotto2020}.
In particular, engineering the band structure is  possible by tuning the hopping rate $J$ between neighboring cavities, which has been shown to lead to nontrivial topological properties~\cite{Wang21, Wang2021, Kim2021, Vega2021, Cheng2021, Besedin2021}.
Moreover, the model could be extended beyond the single-excitation regime to study multiphoton bound states~\cite{Calajo2016, Scigliuzzo2021} and superradiant emission. 
Finally, since the most obvious experimental implementation to date is superconducting qubits, the model could be extended to consider the atoms as three-level systems, to better represent qubits with small anharmonicity.


\begin{acknowledgments}

AS and AFK acknowledge support from the Swedish Research Council (grant number 2019-03696) and from the Knut and Alice Wallenberg Foundation through the Wallenberg Centre for Quantum Technology (WACQT). CSM acknowledges that the project that gave rise to these results received the support of a fellowship from la Caixa Foundation (ID 100010434) and from the European Union’s Horizon 2020 research and innovation programme under the Marie Sk\l{}odowska-Curie grant agreement No 847648, with fellowship code LCF/BQ/PI20/11760026, and financial support from the Proyecto Sin\'ergico CAM 2020 Y2020/TCS-6545 (NanoQuCo-CM).

\end{acknowledgments}


\appendix

\section{Simulating the dynamics with the spectral method}
\label{app:Dynamics}

Here, we explain the algorithm~\cite{Press07, Gonzalez-Tudela2017} we use to simulate the dynamics of the system, as described in \eqref{eq:spectral_method}. For each time step, we proceed in the following way:

\begin{enumerate}
    \item We start with the wave function $\ket{\psi(t_i)}$ written in real space, and apply the evolution of $H_A + \Hint$ which can be precalculated analytically as it is a $2\times2$ Hamiltonian for each coupling point:
    \begin{equation}
        H_A + \Hint = \mqty(\detuning & g\\ g & 0),
    \end{equation}
    with basis states $\ket{e,0}=(1, 0)^T$ and $\ket{g,1}=(0, 1)^T$.
    \item We change the basis of the bath modes to $k$-space by applying a fast Fourier transform algorithm [denoted in \eqref{eq:spectral_method} by $U_{n\to k}$].
    \item We apply the evolution of $H_B$, which can also be precalculated analytically because the Hamiltonian is $N\times N$ diagonal in that representation:
    \begin{equation}
        H_B = -2J\mqty(\dmat[0]{\cos(-\pi), \ddots, \cos(\pi-\frac{2\pi}{N})}).
    \end{equation}
    \item We change the basis again to real space with $U_{k\to n}$ to prepare for the next step. 
\end{enumerate}


\section{Resolvent formalism and derivation of the self-energy}
\label{app:SelfEnergy}

The formalism used here is based on Chapter 3 of Ref.~\cite{Cohen-Tannoudji} and adapted to the particular case of giant atoms with two coupling points.
It is straightforward to generalize to an arbitrary number of coupling points.

\subsection{A single giant atom}
Let us take the total Hamiltonian of a single GA coupled to a structured bath from \eqref{eq:H}, $H=H_0+H_\text{int}$, where
\begin{align}
    H_0 =& \detuning \sigma^+\sigma^- + \sum_k \omega(k) a_k^\dagger a_k\\
    H_\text{int} =& \frac{g}{\sqrt{N}}\sum_k \mleft[\mleft(e^{ikn_1}+e^{ikn_2}\mright)a_k\sigma^+ + \Hc \mright].
\end{align}
In the single-excitation subspace, the eigenstates of the bare Hamiltonian $H_0$ are $\ket{e}:=\ket{e,0}$ and $\ket{k} = \ket{g,k}$, for $k\in [-\pi, \dots, \pi-\frac{2\pi}{N}]$.
The interaction term $H_\text{int}$ couples the subspace $\ket{e}$ with $\ket{k}$.
The \emph{resolvent} of the Hamiltonian is defined by 
\begin{equation}
    G(z) = \frac{1}{z-H}.
\end{equation}
In general, for a projector $P$ of eigenvectors of $H_0$ and its complementary $Q = 1-P$, the resolvent obeys
\begin{equation}
    PG(z)P = \frac{P}{z - PH_0P - P\Sigma(z)P},
\end{equation}
where
\begin{align}
    \Sigma (z) =& H_\text{int} + H_\text{int} \frac{Q}{z-QH_0Q - QH_\text{int}Q}H_\text{int}\nonumber\\
                \approx& H_\text{int} + H_\text{int} \frac{Q}{z-H_0}H_\text{int}
    \label{eq:app_sigma_z}
\end{align}
is the \emph{level-shift operator}.
Note that the approximation symbol above denotes the second-order perturbative expansion in powers of $H_\text{int}$, a truncation that is justified since $H_\text{int}$ is small compared to $H_0$.
In particular, when $P$ is the projector onto a single state $\ket{\alpha}$ with energy $E_\alpha$,
\begin{equation}
    G_\alpha (z) = \frac{1}{z-E_\alpha - \Sigma_\alpha (z)},
    \label{eq:G_alpha}
\end{equation}
with $G_\alpha(z) = \mel{\alpha}{G(z)}{\alpha}$ and $\Sigma_\alpha(z) = \mel{\alpha}{\Sigma(z)}{\alpha}$.

In our case, let us define $P=\ketbra{e}$ and its complementary $Q=\sum_k\ketbra{k}$.
Then we can write the self-energy of the atom $\Sigma_e(z) = \mel{e}{\Sigma(z)}{e}$ as follows:
\begin{align}
    \Sigma_e(z) = &\cancelto{0}{\mel{e}{H_\text{int}}{e}} \quad+ \sum_k \bra{e}H_\text{int}\frac{\ketbra{k}}{z-H_0}H_\text{int}\ket{e} \nonumber\\
    =& \frac{g^2}{N}\sum_k \frac{(e^{ikn_1}+e^{ikn_2})(e^{-ikn_1}+e^{-ikn_2})}{z-\omega(k)}\nonumber\\
    =&\frac{2g^2}{N}\sum_k \frac{1+\cos(k(n_2-n_1))}{z-\omega(k)}.
    \label{eq:app_sigma_cos}
\end{align}
Henceforth, we use the dispersion relation $\omega(k)=-2J\cos(k)$ and that the distance between coupling points is $d=n_2-n_1$.
In the continuum limit, i.e., when $N\to \infty$, the sum over $k$ becomes an integral: $\sum_k\frac{2\pi}{N}\to \int_k dk.$
Therefore, we can write the self-energy like
\begin{align}
    \Sigma_e(z)
    =& \frac{g^2}{\pi}\int_{-\pi}^\pi \frac{1+\cos(kd)}{z+2J\cos(k)}dk \nonumber\\
    =&  \frac{g^2}{\pi}\int_{-\pi}^\pi \frac{dk}{z+2J\cos(k)} +  \frac{g^2}{\pi}\int_{-\pi}^\pi \frac{e^{ikd}}{z+2J\cos(k)}dk,
\end{align}
where, in the second integral, we have substituted the cosine for an exponential because odd functions do not contribute to the integral.
Now, we can introduce the change of variable $\tilde{z} = e^{ik}$ such that $2\cos(k)=\tilde{z}+\tilde{z}^{-1}$ and $dk=-i\tilde{z}^{-1}d\tilde{z}$, and integrate over the unit circle:
\begin{align}
    \Sigma_e(z)
    =&  -\frac{ig^2}{\pi}\oint \frac{d\tilde{z}}{z\tilde{z}+J\tilde{z}^2 + 1} - \frac{ig^2}{\pi}\oint \frac{\tilde{z}^d d\tilde{z}}{z\tilde{z}+J\tilde{z}^2 + 1}\nonumber\\
    =&-\frac{ig^2}{\pi}\oint \frac{(1+\tilde{z}^d)d\tilde{z}}{(\tilde{z}-f_+)(\tilde{z}-f_-)},
\end{align}
where $f_\pm(z)= (-z\pm \sqrt{z^2-4J^2})/(2J)$ are the roots of the quadratic equation, as defined in \eqref{eq:f(z)}.
Applying the residue theorem, we obtain that
\begin{equation}
    - \frac{ig^2}{\pi}\oint \frac{d\tilde{z}}{(\tilde{z}-f_+)(\tilde{z}-f_-)} = \frac{2g^2}{\sqrt{z^2-4J^2}},
\end{equation}
and thus the final expression of the self-energy becomes
\begin{equation}
    \Sigma_e(z) = \pm \frac{2g^2}{\sqrt{z^2-4J^2}} \mleft[1+\mleft(\frac{-z\pm \sqrt{z^2-4J^2}}{2J}\mright)^d\mright].
    \label{eq:app_Sigma_e}
\end{equation}
Note that this expression is the particular case of \eqref{eq:SelfEnergy1} for a GA with two coupling points.

According to \eqref{eq:G_alpha}, the resolvent operator element corresponding to the excited state of the atom is then
\begin{equation}
    G_e(z) = \frac{1}{z-\detuning-\Sigma_e(z)},
\end{equation}
with $\detuning$ being the atom-cavity detuning.
Lastly, we can express the probability amplitude of an initially excited GA, for $t>0$, as follows:
\begin{equation}
    C_e(t) = -\frac{1}{2\pi i}\int_{-\infty}^{\infty} dE e^{-iEt} G_e(E + i 0^+).
\end{equation}

\subsection{Two giant atoms}

For two identical giant atoms, we can consider the projector $P=\ketbra{eg} + \ketbra{ge}$, which refers to one atom being excited and the other in the ground state.
In such a case, the matrix representing $PG(z)P$ in the basis $\{\ket{eg}, \ket{ge}\}$ is thus a $2\times2$ matrix:
\begin{equation}
    \mqty(G_{eg} & G_{eg, ge}\\ G_{ge, eg} & G_{ge}) = \mqty(z-\detuning - \Sigma_{eg} & -\Sigma_{eg, ge}\\
    -\Sigma_{ge, eg} & z-\detuning-\Sigma_{ge})^{-1},
\end{equation}
where $\detuning$ denotes the atom-cavity detuning for each atom and the diagonal terms have been written with simplified subscripts (e.g., $G_{eg}= G_{eg,eg} = \mel{eg}{G(z)}{eg}$ and $\Sigma_{ge}= \Sigma_{ge,ge} = \mel{ge}{\Sigma(z)}{ge}$).
Since the atoms are identical, $\Sigma_{eg} =\Sigma_{ge} = \Sigma_{e}$, with $\Sigma_e$ from \eqref{eq:app_Sigma_e}, and $\Sigma_{eg,ge}=\Sigma_{ge,eg}=\Sigma_\text{int}$, with $\Sigma_\text{int}$ from \eqref{eq:Sigma_int}.

It is, however, more convenient to work in the basis spanned by $\ket{\pm}=(\ket{eg}\pm\ket{ge})/\sqrt{2}$, where
\begin{equation}
    \mqty(G_{+} & G_{+, -}\\ G_{-, +} & G_{-}) = \mqty(z-\detuning - \Sigma_{+} & 0\\
   0 & z-\detuning-\Sigma_{-})^{-1},
\end{equation}
since $G$ is diagonal and $\Sigma_{\pm} = \Sigma_{e}\pm\Sigma_{\text{int}}$.
In this basis, the probability amplitude associated to the states $\ket{\pm}$ is, as shown in \eqref{eq:C_pm},
\begin{align}
    C_\pm(t) =&  -\frac{1}{2\pi i}\int_{-\infty}^\infty G_\pm(E + i0^+)e^{-iEt} dE \nonumber\\
    =& -\frac{1}{2\pi i}\int_{-\infty}^\infty \frac{e^{-iEt} dE}{(E+i0^+) -\detuning-(\Sigma_{e}\pm\Sigma_{\text{int}})}.
\end{align}
We can then go back to the basis $\{\ket{eg}, \ket{ge}\}$ and calculate the probability amplitude of starting with one of the atoms in the excited state. For instance, when the initial state is $\ket{eg}$:
\begin{align}
    C_{eg}(t) =& -\frac{1}{2\pi i}\int_{-\infty}^\infty \underbrace{G_{eg}(E + i0^+)}_{\mel{eg}{G(E+i0^+)}{eg}}e^{-iEt} dE \nonumber\\
    =& -\frac{1}{2\pi i}\int_{-\infty}^\infty \frac{1}{2}\Big(\underbrace{\mel{+}{G}{+}}_{G_+} + \cancelto{0}{\mel{-}{G}{+}} \nonumber\\
    &\qquad\qquad\quad + \cancelto{0}{\mel{+}{G}{-}} + \underbrace{\mel{-}{G}{-}}_{G_-}\Big) e^{-iEt} dE \nonumber\\
    =&\frac{1}{2}[C_+(t)+ C_-(t)].
\end{align}
Similarly, for the initial state $\ket{ge}$,
\begin{equation}
    C_{ge}(t) = \frac{1}{2}[C_+(t)- C_-(t)].
\end{equation}


\bibliography{References.bib}

\end{document}